\documentclass[10pt]{iopart}
\usepackage{graphicx}

\begin{document}

\title{Eliashberg analysis of the optical conductivity in superconducting Pr$_2$CuO$_{x}$ ($x \simeq 4$)}

\author{E Schachinger$^{1}$, G Chanda$^{2,3}$, R P S M Lobo$^{4,5,6}$,
M Naito$^{7}$ and A V Pronin$^{2,8}$}
\address{$^{1}$ Institute of Theoretical and Computational Physics,
NAWI Graz, Graz University of Technology, A-8010 Graz, Austria}

\address{$^{2}$ Dresden High Magnetic Field Laboratory (HLD),
Helmholtz-Zentrum Dresden-Rossendorf, 01314 Dresden, Germany}

\address{$^{3}$ Institut f\"{u}r Festk\"{o}rperphysik, Technische
Universit\"{a}t Dresden, 01062 Dresden, Germany}

\address{$^{4}$ LPEM, PSL Research University, ESPCI ParisTech, 10
rue Vauquelin, 75231 Paris Cedex 5, France}

\address{$^{5}$ CNRS, UMR8213, Paris, France}

\address{$^{6}$ Sorbonne Universit\'es, UPMC Univ. Paris 6, 75005, Paris, France}

\address{$^{7}$ Department of Applied Physics, Tokyo University of
Agriculture and Technology, Naka-cho 2-24-16, Koganei, Tokyo
184-8588, Japan}

\address{$^{8}$ A. M. Prokhorov Institute of General Physics, Russia Academy of Sciences,
119991 Moscow, Russia}

\eads{ewald.schachinger@live.at, pronin@ran.gpi.ru}

\date{\today}

\begin{abstract}
Superconducting Pr$_2$CuO$_x$, $x\simeq 4$ films with $T^\prime$
structure and a $T_c$ of 27 K have been investigated by
millimeter-wave transmission and broadband
(infrared-to-ultraviolet) reflectivity measurements in the normal
and superconducting state. The results obtained by both
experimental methods show a consistent picture of the
superconducting condensate formation below $T_c$. An Eliashberg
analysis of the data proves $d$-wave superconductivity and
unitary-limit impurity scattering of the charge carriers below
$T_{c}$. The derived electron-exchange boson interaction spectral
function $I^2\chi(\omega)$ shows only marginal changes at the
superconducting transition with the mass enhancement factor
$\lambda$, the first inverse moment of $I^2\chi(\omega)$, being
equal to 4.16 at 30 K and to 4.25 at 4 K.
\end{abstract}


\section{Introduction}

The parent compounds of the superconducting high-$T_c$ cuprates
are generally considered to be antiferromagnetic charge-transfer
insulators (CTIs) \cite{Imada, ARM}. For the hole-doped compounds,
this statement is commonly accepted. For the electron-doped
cuprates, the situation is trickier: there are studies, that put a
question mark on this picture.

For example, in 1995 Brinkmann~\textit{et al.}~\cite{Brinkmann}
demonstrated that the superconducting phase in
Pr$_{2-x}$Ce$_x$CuO$_4$ (PCCO) single crystals can exist at doping
levels as low as 4\,\%. This was achieved by a special
oxygen-reduction and annealing technique. A large body of work
performed in the last few years on thin films demonstrated that
even the parent compounds of the electron-doped cuprates, e.g.
(La,Y)$_2$CuO$_{4}$ and $R_2$CuO$_4$ with $R$ being Pr, Sm, Nd,
Eu, or Gd, are metallic and become superconducting at low
temperatures \cite{Matsumoto1, Matsumoto2, Matsumoto3, Matsumoto4,
Yamamoto, Ikeda, YKrock1, Kojima}. Finally, superconductivity was
reported in nominally undoped polycrystalline samples of
(La,Sm)$_2$CuO$_{4}$~\cite{Ueda, Asai} and in heavily underdoped
single crystals of the Pr-La-Ce-Cu-O system \cite{TAdachi}.

\begin{figure}[t]
\centering
\includegraphics[width=10 cm,clip]{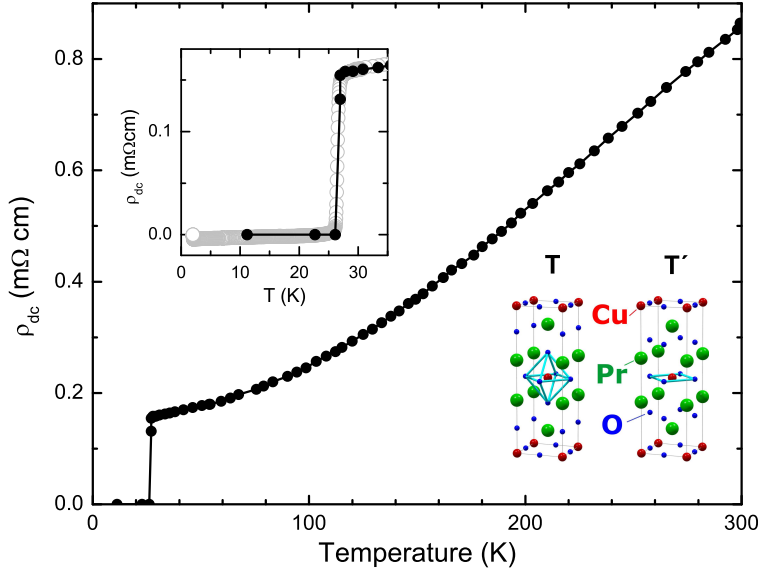}
\caption{(Color online) Main panel: Temperature dependence of the
in-plane dc resistivity $\rho_{dc}$ of the $T^\prime$-PCO film. In
the right-hand corner schematic diagrams illustrate the $T$ and
$T^\prime$ structures. Inset: A close-up of the dc resistivity
around the superconducting transition. The dc measurements were
performed twice: on the fresh film [solid (black) symbols] and
after completion of all optical measurements [open (grey)
symbols].} \label{rho}
\end{figure}

The sharp contradiction between these and earlier results can be
explained as being due to structure-related issues, namely due to
the presence or absence of oxygen atoms in apical positions, i.e.
in positions directly above and below the copper atoms in the
CuO$_{2}$ planes, as illustrated in the right-hand corner of
Figure~\ref{rho}. While the hole-doped cuprates have the so-called
$T$ structure characterized by the presence of apical oxygen, the
electron-doped cuprates are supposed to posses the $T^\prime$
structure, with no apical oxygen. In practice, though, it is very
hard to remove the apical oxygen from the electron-doped cuprates
completely, particularly in bulk samples \cite{YKrock1}. Apical
oxygen in the $T^\prime$ structure is known to act as a very
strong scatterer and pair breaker \cite{Sekitani} and, thus, the
$T_{c}$ reduction in the underdoped compounds may in fact be due
to the remaining apical oxygen. Because of tenuity of films and
their large surface-to-volume ratio, the process of removing extra
oxygen (oxygen reduction) is easier to perform and control in
films rather than in bulk samples. This makes films advantageous
for achieving the proper $T^\prime$ structure with no apical
oxygen.

Superconductivity reported in undoped cuprates challenges the
applicability of the CTI picture to electron-doped cuprates
\cite{Naito}. It is worth noting here that recent calculations
performed on the basis of a new first-principles method report a
sharp difference between the parent compounds with $T$ and
$T^\prime$ structures \cite{Das, Weber1, Weber3}. While the first
are found to be ``standard" CTIs, the latter, e.g.
Pr$_{2}$CuO$_{4}$, are essentially metallic and their apparent
insulating nature may originate from magnetic long-range order
(Slater transition) which is competing with the metallic ground
state.

However, it is still an open question whether the supposedly
undoped $T^\prime$ superconductors are truly undoped. An
alternative possibility is a doping by oxygen vacancies in the
$R$O layers. These vacancies may in principle be created during
the oxygen-reduction process. Because no single-crystalline bulk
$T^\prime$-$R_2$CuO$_{4}$ superconducting samples have yet been
synthesized, direct measurements of the oxygen distribution are
impossible so far. Nevertheless, neutron diffraction experiments
performed on Nd$_{2-x}$Ce$_{x}$CuO$_{4+y}$ single crystals showed
that it was mostly apical oxygen, which was removed during the
oxygen reduction \cite{Schultz, Radaelli}. The recent report on
superconductivity in bulk heavily underdoped
$T^\prime$-Pr$_{1.3-x}$La$_{0.7}$Ce$_{x}$CuO$_{4+\delta}$ gives
hope that the oxygen distribution issue for the $T^\prime$
cuprates might be clarified in the near future \cite{TAdachi}.

\begin{figure}[t]
\centering
\includegraphics[width=10 cm,clip]{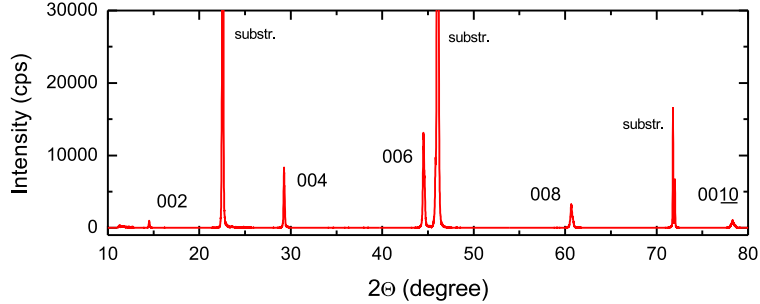}
\caption{(Color online) X-ray diffraction pattern of a $c$-axes
oriented PCO film on DyScO$_{3}$ substrate. Diffraction peaks
coming from the substrate are marked as ``substr." The lattice
constant along the $c$-axes calculated from the positions of the
film peaks is 12.20 \AA, indicating the optimal oxygen reduction
and the proper $T^\prime$ structure \cite{Matsumoto1}.}
\label{x_ray}
\end{figure}

In our recent paper \cite{Chanda} we investigated the optical
response of MBE-grown Pr$_{2}$CuO$_{x}$ (PCO) films with $x \simeq
4$. We showed that the optical response can be consistently
understood within the picture in which superconductivity develops
in undoped PCO, i.e. $x$ is indeed 4. (Although it is quite
obvious that one cannot judge whether or not the CTI picture is
valid from optical measurements alone). In that paper, we reported
on \textit{normal-state} broadband optical reflectivity
measurements and on millimeter-wave measurements below the
superconducting transition. We demonstrated that neither raw
experimental optical data nor their analysis reveal any indication
of normal-state gap-like features which could be attributed to the
existence of a normal-state pseudogap. This observation is in line
with the breakdown of the CTI picture in PCO. In our
millimeter-wave measurements, we directly observed the formation
of the superconducting condensate at $T < T_{c}$ and found that
the temperature dependence of the London penetration depth at low
temperatures follows a quadratic power law indicating a $d$-wave
pairing symmetry typical for cuprate superconductors. Similar
results on penetration depth for PCO films, prepared by
metal–organic decomposition, were also reported by some of us a
few years earlier \cite{Pronin}.

In this paper, we report on the broadband optical conductivity
\textit{in the superconducting state}, measured on the same sample
as in Ref.~\cite{Chanda}. We perform Eliashberg analysis of
optical data collected below $T_{c}$ and compare the results to
the normal-state data of Ref.~\cite{Chanda}. We demonstrate that
the results obtained by both our techniques (broadband optical
reflectivity and millimeter-wave transmission measurements) for
the superconducting state reveal a consistent picture of the
superconducting condensate formation. We find that $d$-wave
superconductivity and unitary-limit electron scattering off
impurities describe best the available experimental data below
$T_{c}$.

\section{Samples and Experiment}

As described in Ref.~\cite{Chanda}, the PCO films were grown by
molecular beam epitaxy (MBE) \cite{Yamamoto} on a (110)-oriented
0.35 mm thick DyScO$_{3}$ substrate. The films were 100 nm thick
with the $c$-axis oriented perpendicular to the film's surface. We
investigated two thin films of PCO. The results, obtained on the
films, do not demonstrate any significant difference. Hereafter we
discuss results gathered from one of the two films, the same one
as in Ref.~\cite{Chanda}.

Figures \ref{rho} and \ref{x_ray} show the results of dc
resistivity and x-ray diffraction measurements performed on the
film. The sharp diffraction peaks with no signs of spurious phases
demonstrate the film's quality. The resistivity decreases
monotonically with decreasing temperature down to $T_c = 27$ K.
The width of the superconducting transition is 0.8 K. The $T_c$
and the transition width remained unchanged after completion of
all optical measurements (as evident from the inset of
Figure~\ref{rho}) indicating that the film did not degrade in the
course of our measurements.

We measured near-normal reflectivity from 40 to 55000 cm$^{-1}$ at
selected temperatures from 4 to 300 K and phase-sensitive
transmission at 210 and 250 GHz (7 and 8.3 cm$^{-1}$) as a
function of temperature, using backward wave oscillators (BWOs).
Details of these measurements have been discussed by us earlier
\cite{Chanda}.

\begin{figure}[b]
\centering
\includegraphics[width=10 cm,clip]{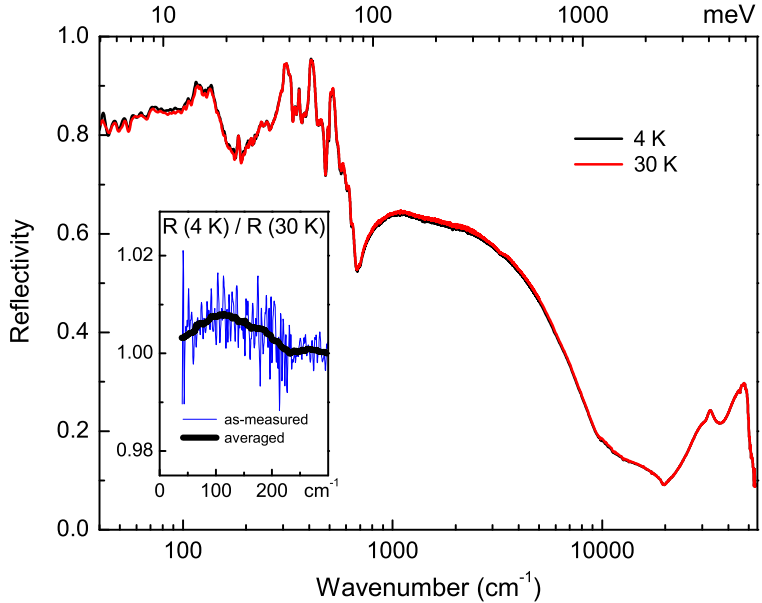}
\caption{(Color online) Reflectivity of the PCO thin film on a
DyScO$_{3}$ substrate as a function of frequency at 4 and 30 K.
The \textbf{E} vector of the probing radiation lies in the
$ab$-plane of the film (and parallel to the [001] axis of the
substrate). The inset shows the superconducting-state (4 K)
reflectivity divided by the normal-state (30 K) reflectivity at
the lowest frequencies. The thin (blue) line corresponds to the
measured and the thick (black) line to the frequency averaged
spectrum.} \label{reflectivity}
\end{figure}

\section{Experimental results}

Figure~\ref{reflectivity} shows the in-plane ($ab$-plane)
reflectivity of the PCO film on a DyScO$_{3}$ substrate versus
frequency at $4\,$K and $30\,$K, e.g. at the lowest possible $T$
in the superconducting state and at a temperature slightly above
the $T_{c}$. At low frequencies, the reflectivity is quite high,
typical for metals and metallic films. At 100 -- 700 cm$^{-1}$, a
number of phonon modes from both, the substrate and the film,
affect the spectra. From 1000 to 10000 cm$^{-1}$, the reflectivity
monotonously decreases with frequency. The maxima seen above 10000
cm$^{-1}$ are due to inter-band transitions \cite{Chanda}.

The inset of Figure~\ref{reflectivity} discusses the ratio
$R_{s}/R_{n}$ of the reflectivity at $4\,$K ($R_{s}$) and $30\,$K
($R_{n}$). Changes in the spectrum induced by the superconducting
transition are not very pronounced but, nevertheless, can clearly
be seen at frequencies below 200 cm$^{-1}$. If we reduce the
frequency resolution [thick (black) line] the changes become more
apparent. Let us emphasize, though, that because of the fact that
the film is thin and partly transparent (the absolute reflectivity
at low frequencies is only about 0.8) pronounced changes in the
reflectivity induced by superconductivity cannot be expected in
contrast to what can be observed in bulk samples.

For conventional isotropic $s$-wave superconductors the maximum in
$R_{s}/R_{n}$ is known to indicate the position of the
superconducting energy gap $2\Delta_0$ \cite{Tinkham}. For a rough
estimate of $2\Delta_0$ such an approach can also be applied to
the cuprates. If we associate in a straightforward way the maximum
in $R_{s}/R_{n}$ to the energy gap, then we obtain $2\Delta_0$ =
110 cm$^{-1}$ and $2\Delta_0/k_{B}T_{c}$ = 5.8. This value is
roughly in the middle of the broad range of $2\Delta_0/k_{B}T_{c}$
obtained by different experimental methods for the electron-doped
cuprates \cite{ARM} and is close to the results of optical
measurements on PCCO with different doping levels \cite{Zimmers1,
Homes2}.

The film's complex optical conductivity, $\sigma = \sigma_{1} +
i\sigma_{2}$, has been extracted from the reflectivity spectra by
use of a thin-film Drude-Lorentz fitting procedure which was
described in necessary detail by Chanda \textit{et al.}
\cite{Chanda}. This procedure is similar to the one proposed by
Kuzmenko \cite{Kuzmenko} and can be as accurate as the
Kramers-Kronig analysis (which is hardly possible for
partly-transparent films on substrates). Neither BWO data nor
values of the dc conductivity in the normal state have been
utilized within this fitting procedure.

\begin{figure}[t]
\centering
\includegraphics[width=10 cm,clip]{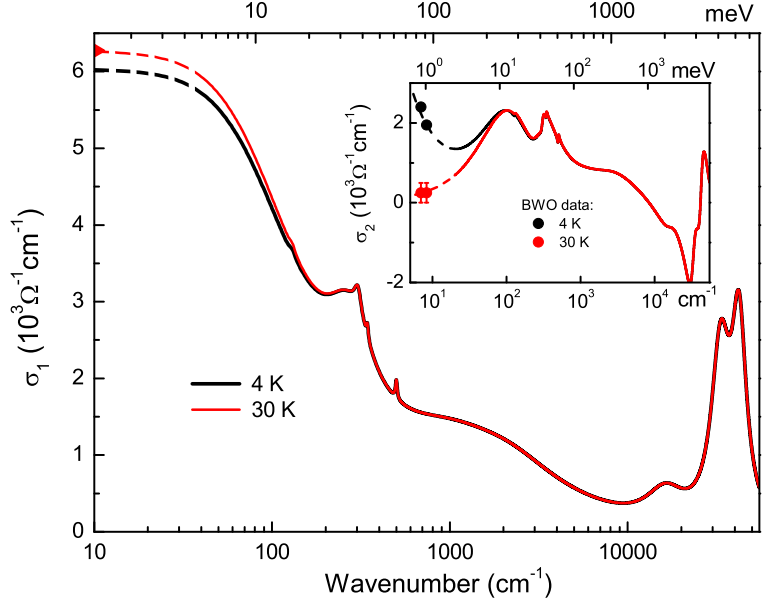}
\caption{(Color online) Real (main panel) and imaginary (inset)
part of the optical conductivity of PCO as a function of frequency
for 4 [(black) lines] and 30 K [(red) lines]. The (red) right-hand
triangle on the left-hand axis of the main panel represents the
dc-conductivity at 30 K. Solid circles in the inset indicate
$\sigma_{2}(\omega)$ data as obtained from the BWO phase-sensitive
transmission measurements at 7 and 8.3 cm$^{-1}$. Error bars for
the $T=4\,$K BWO data are within symbol size. The dashed lines
starting below 40 cm$^{-1}$ correspond to extrapolations obtained
from the Drude-Lorentz fitting procedure as described by Chanda
\textit{et al.} \cite{Chanda}.} \label{conductivity}
\end{figure}

The real part of the PCO optical conductivity obtained by this
modeling at $4\,$K and $30\,$K is presented in
Figure~\ref{conductivity}. The lowest frequency of the
reflectivity measurements was 40 cm$^{-1}$. Therefore, the data
obtained from this analysis below this threshold frequency is to
be considered as an extrapolation and is indicated by dashed lines
as a guide to the eye. Nevertheless, the zero-frequency limit of
$\sigma_{1}$ at $30\,$K evolves in accordance with $\sigma_{dc}$
[(red) triangle on the vertical left-hand axis of
Figure~\ref{conductivity}]. At low frequencies, the
superconducting-state $\sigma_1(\omega)$ is below its normal state
values. However, a large Drude-like contribution persists in the
superconducting state. Similar behavior of optical conductivity
has also been reported e.g. for La$_2$Sr$_{2-x}$CuO$_4$
\cite{Gorshunov} and Bi$_2$Sr$_2$CaCu$_2$O$_{8+\delta}$
\cite{Santander, Corson}.

The inset of Figure~\ref{conductivity} presents the data of the
imaginary part of the optical conductivity, $\sigma_2(\omega)$, at
$4\,$K [(black) line] and $30\,$K [(red) line]. In addition, the
independent BWO data at 7 and 8.3 cm$^{-1}$ are indicated by solid
circles and it is evident that at $T=4 \,$K $\sigma_2(\omega)$
tends to diverge for $\omega\to 0$, while at $T=30\,$K
$\sigma_2(\omega)$ tends to zero within the same limit. Thus, our
data for the superconducting and normal state are not only
distinct from each other but they also exhibit the expected
behavior \cite{Tinkham} in the superconducting state: a divergence
at $\omega\to 0$, and in the normal state: a Drude metal with
diminishing $\sigma_2(\omega)$ at $\omega\to 0$.

\section{Data Analysis}

We start our analysis with the so-called extended Drude model
\cite{JWAllen, Puchkov}. In this model the complex conductivity is
given by
\begin{equation}
\sigma(\omega) =
\frac{1}{4\pi}\frac{\omega_{p}^{2}}{\Gamma(\omega)
-i\omega[1+\lambda(\omega)]}, \label{ext_Drude1}
\end{equation}
where $[1+\lambda(\omega)] = m^{*}(\omega)/m$ and
$\tau_{op}^{-1}(\omega) \equiv\Gamma(\omega)$ are the
frequency-dependent mass renormalization factor and the optical
scattering rate, respectively. Inverting equation
(\ref{ext_Drude1}) gives:
\begin{equation}
1+\lambda(\omega) =
\frac{\omega_{p}^{2}}{4\pi}\frac{\sigma_{2}(\omega)}{\omega|\sigma(\omega)|^{2}};
\quad \Gamma(\omega) =
\frac{\omega_{p}^{2}}{4\pi}\frac{\sigma_{1}(\omega)}{|\sigma(\omega)|^{2}}.
\label{ext_Drude2}
\end{equation}

\begin{figure}[t]
  \centering
  \includegraphics[width=10 cm,clip]{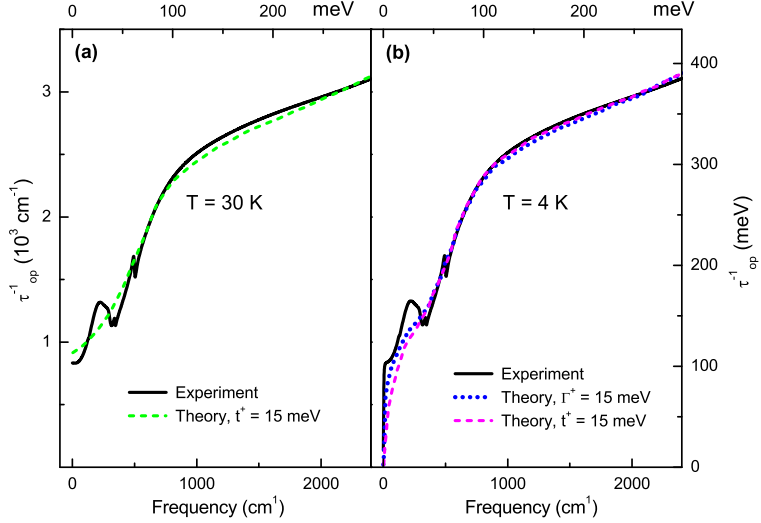}
  \caption{(Color online) The optical scattering rate $\tau^{-1}_{op}(\omega)$
vs energy $\omega$. Frame (a): Normal state results at $T=30\,$K.
The solid (black) curve represents the data while the dashed
(green) line corresponds to the theoretical result of normal-state
Eliashberg theory using the $I^2\chi(\omega)$ spectrum presented
in Figure~\ref{fig:I2c} by a dashed (black) line. The impurity
parameter $t^+=15\,$meV. Frame (b): Superconducting state results
at $T=4\,$K. The solid (black) curve represents the data. The
dotted (blue) line corresponds to results generated by $d$-wave
symmetry Eliashberg theory on the basis of the $I^2\chi(\omega)$
spectrum presented in Figure~\ref{fig:I2c} [solid (red) curve].
Impurity scattering was described in the unitary limit, $\Gamma^+
= 15\,$meV. The dashed (magenta) curve presents the equivalent
result for Born limit scattering with $t^+ = 15\,$meV. Definitions
of $t^+$ and $\Gamma^+$ are given in the text.}
  \label{fig:InvTau}
\end{figure}

The frequency-dependent optical scattering rate, as obtained from
our data using (\ref{ext_Drude2}) with $\omega_{p}$ = 2.19 eV
\cite{Chanda}, is displayed in Figures~\ref{fig:InvTau}(a,b) as a
function of frequency for $30\,$K and $4\,$K [solid (black)
curves], respectively. The general trend in $\tau_{op}^{-1}
(\omega)$ is to increase with frequency, but this increase is
non-monotonic. This is due to phonons and a localization mode seen
as a bump at around 230 cm$^{-1}$ as discussed by Chanda
\textit{et al.} \cite{Chanda}.

There is only little difference between the $4\,$K and $30\,$K
$\tau^{-1}_{op}(\omega)$ data. This is quite similar to what has
been reported for PCCO by Schachinger {\it et al.} \cite{ESchach}.
Nevertheless, it is important to emphasize that according to
(\ref{ext_Drude2}) and the results for $\sigma_2(\omega)$ (see
inset of Figure~\ref{conductivity}) the normal-state
$\tau^{-1}_{op}(\omega)$ stays finite in the limit $\omega\to0$.
On the other hand, it drops precipitately to zero in the rather
narrow energy range of $\omega\in[0,\sim50]\,$cm$^{-1}$ at
$T=4\,$K. To emphasize this feature we added a small negative
offset to the frequency-scale in Figures~\ref{fig:InvTau}(a,b).

The difference between total optical scattering rate and impurity
scattering rate, i.e. $\tau^{-1}_{op}(\omega) - \tau^{-1}_{imp}$,
is according to P. B. Allen \cite{allen} closely related to the
electron-exchange boson interaction spectral density
$I^2\chi(\omega)$ which is at the core of the normal and
superconducting state Eliashberg theory \cite{ESchach3}. Thus, it
is of quite some interest to gain knowledge on $I^2\chi(\omega)$
by inverting this difference using methods which have been
discussed in detail by Schachinger \textit{et al.}
\cite{ESchach2}. It was, furthermore, demonstrated by Schachinger
\textit{et al.} \cite{ESchach1} that any non-zero contribution to
$I^2\chi(\omega)$ at some energy $\omega$ will always result in an
increase of the optical scattering rate. Consequently the bump
observed in the experimental optical scattering rate of PCO
(Figure~\ref{fig:InvTau}) at around 230 cm$^{-1}$ ($\sim28\,$meV)
cannot be caused by electron-exchange boson interaction and is,
therefore, not part of the conducting-electron background.

\begin{figure}[b]
  \centering
  \includegraphics[width=10 cm,clip]{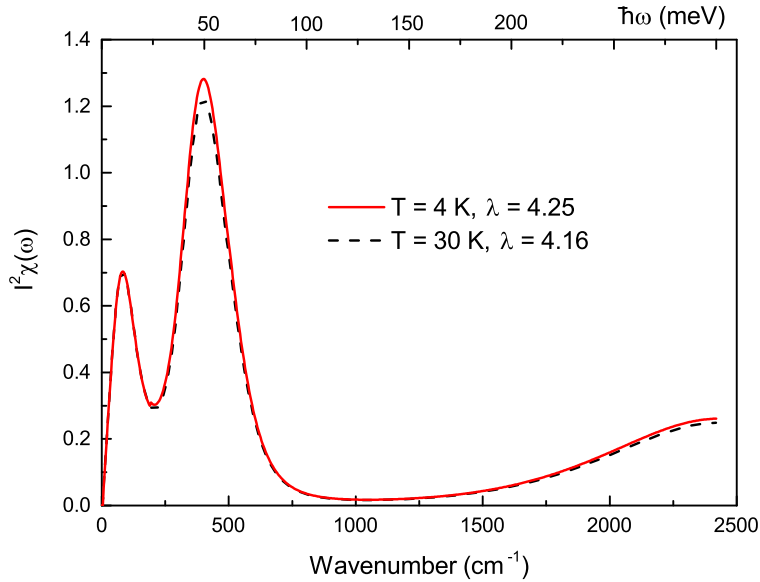}
  \caption{(Color online) The $I^2\chi(\omega)$ spectrum for two
temperatures, namely $T=4\,$K [solid (red) line] and $30\,$K
[dashed (black) line] as a result of a straightforward inversion
of the experimental $\tau^{-1}_{op}(\omega)$ data shown in
Figure~\ref{fig:InvTau}a. $\lambda$ is the first inverse moment of
$I^2\chi(\omega)$ or the mass enhancement factor.}
  \label{fig:I2c}
\end{figure}

The inversion of the normal state $T=30\,$K data using a maximum
entropy method of inversion \cite{ESchach1} has already been
discussed in quite some detail by Chanda \textit{et al.}
\cite{Chanda} and resulted in an $I^2\chi(\omega)$ spectrum
featuring a double peak structure followed by a deep valley and a
hump at higher energies. We reproduce these results for comparison
in Figure~\ref{fig:InvTau}(a) which shows the experimental data by
a solid (black) line and the data reconstruction due to
normal-state Eliashberg theory by a dashed (green) line. The
resulting $I^2\chi(\omega)$ is presented by a dashed (black) line
in Figure~\ref{fig:I2c}. The low-energy peak is at $\sim 11\,$meV
and the high energy peak can be found at $\sim 50\,$meV. This
inversion process resulted also in an impurity scattering rate of
$\tau^{-1}_{imp} = 2\pi t^+ = 100\,$meV ($t^+ = 15\,$meV) which is
quite substantial but in reasonable agreement with what has been
reported for the system PCCO \cite{ESchach}. Similar double peak
spectra have been reported for PCCO by Schachinger {\it et al.}
\cite{ESchach} and for La$_{1.83}$Sr$_{0.17}$CuO$_4$ (a hole doped
cuprate) by Hwang {\it et al.} \cite{Hwang} both with a less
pronounced low-energy peak. It is most likely that the bump around
$\sim 28\,$meV in the PCO $\tau^{-1}_{op}(\omega)$ data is
responsible that the low-energy peak is too pronounced in the PCO
$I^2\chi(\omega)$ spectrum.

\begin{figure}[t]
\centering
\includegraphics[width=\columnwidth,clip]{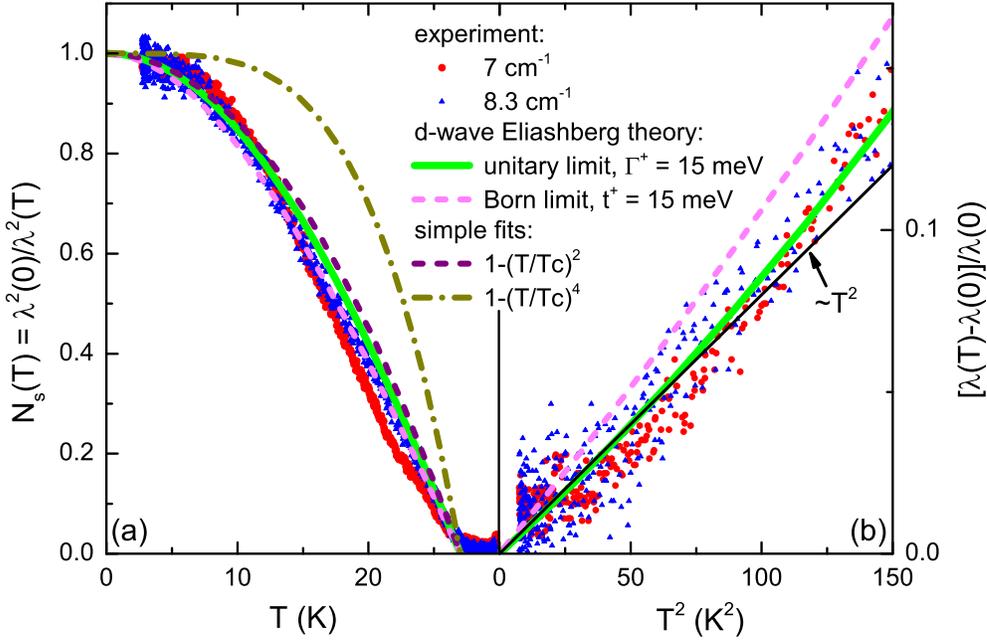}
\caption{(Color online). Frame (a): Superfluid density, $N_s(T) =
\lambda_L^2(0)/\lambda_L^2(T)$, as a function of temperature.
Frame (b): Low-temperature variation of the normalized London
penetration depth, $[\lambda_L(T)-\lambda_L(0)]/\lambda_L(0)$, as
a function of temperature squared. Data derived from the
millimeter-wave conductivity measurements at 7 cm$^{-1}$ and 8.3
cm$^{-1}$ are presented by solid (red) circles and solid (blue)
triangles, respectively. Theoretical results generated by a full
$d$-wave Eliashberg-theory calculation are shown for unitary
scattering [thick solid (green) line] and Born limit scattering
[think dashed (magenta) line]. Frame (a) contains for comparison
the temperature dependence $N_s(T) = 1-(T/T_c)^2$ [thin dashed
(purple) line] which is expected for a nodal superconductor and
$N_s=1-(T/T_c)^4$ [thin dashed-dotted (olive) line] for a fully
gaped superconductor. In Frame (b) a quadratic power law of the
reduced penetration depth is indicated by a thin solid (black)
line.} \label{lambda}
\end{figure}

More information on $I^2\chi(\omega)$ can be gained from an
inversion of superconducting state data together with a comparison
between data and theory. An inversion of the $T=4\,$K
superconducting state optical scattering rate data should be
possible \cite{ESchach2}, but the formula quoted there is only
valid in the clean-limit case and cannot be applied here. The only
possibility that remains is to use the $I^2\chi(\omega)$ spectrum
for $T=30\,$K as a starting point and to calculate the optical
scattering rate in the superconducting state at $T=4\,$K using the
full $d$-wave Eliashberg theory \cite{ESchach3}. The use of
$d$-wave Eliashberg theory is justified by the fact that neither
the superconducting-state data of $\sigma_1(\omega)$ nor the
optical scattering rate show the signature of a full $s$-wave gap,
where $\sigma_1(\omega)$ and $\tau^{-1}_{op}(\omega)$ would be
equal to zero within the energy range $\omega\in[0,2\Delta_0]$
with $\Delta_0$ the superconducting gap. Moreover, the use of the
normal state $I^2\chi(\omega)$ as a staring point is also
justified by the fact that there is very little difference between
the $T=30\,$K and $4\,$K data except, of course, in the vicinity
of $\omega=0$. Thus, a least-squares fit procedure was employed to
generate the superconducting state $I^2\chi(\omega)$ spectrum
which allowed an optimum fit to the data. This spectrum is shown
in Figure~\ref{fig:I2c} by a solid (red) curve and is only
marginally different from the corresponding normal-state derived
spectrum. The mass enhancement factor $\lambda$ also varies only
little from 4.16 at $30\,$K to 4.25 at $4\,$K.

Theory is compared with experiment in Figure~\ref{fig:InvTau}(b).
In contrast to the normal state, the scattering of the electrons
off impurities is in the superconducting state more accurately
described by a $T$-matrix approximation. Here, the impurity
scattering rate is described by a parameter $\Gamma = 2\pi\Gamma^+
= n_I/[N(0)\pi^2]$ with $n_I$ the concentration of scattering
centers and $N(0)$ is the electronic density of states at the
Fermi level. In addition, there is a scattering phase shift
$\delta_0$ and we introduce an additional parameter
$c=\rm{cot}(\delta_0)$ for convenience. The case $c=0$ corresponds
to unitary scattering and for $\delta_0\to 0$, i.e. $c\to\infty$,
the weak scattering limit (Born's approximation) is recovered
\cite{prohammer}. (In the normal state only the weak scattering
limit matters.) Consequently, Figure~\ref{fig:InvTau}(b) presents
results for unitary [dotted (blue) line] as well as Born limit
scattering [dashed (magenta) line] off impurities. As in the
normal state there is little agreement between theory and
experiment in the energy range around the bump at $\sim28\,$meV.
Nevertheless, it is obvious that theory with unitary scattering
describes rather well the precipitous drop of
$\tau^{-1}_{op}(\omega)$ to zero as $\omega\to0$, while in the
case of Born-limit scattering the decrease of
$\tau^{-1}_{op}(\omega)$ as $\omega\to0$ is more moderate.

If we want to discriminate even better between unitary and Born
limit scattering we are required to analyze other experimental
data which are linearly independent from the optical scattering
rate data. One possibility is offered by the London penetration
depth, $\lambda_L(T)$, derived from our phase-sensitive
millimeter-wave transmission measurements \cite{Chanda}. From the
theoretical point of view, $\lambda_L(T)$ can easily be calculated
from $d$-wave Eliashberg theory using the $T=4\,$K spectrum
presented in Figure~\ref{fig:I2c} together with the relevant
impurity parameters because there is only little variation in the
spectrum in going from $4\,$K to $30\,$K.

In Figure~\ref{lambda}(a) we replot the temperature dependence of
the normalized superfluid density, $N_s(T) = n_s(T)/n_s(0) =
\lambda_L^2(0)/\lambda_L^2(T)$, which was obtained from
millimeter-wave measurements at 7 cm$^{-1}$ and 8.3 cm$^{-1}$ and
reported by us in Ref.~\cite{Chanda}. Curves for $N_s(T) =
1-(T/T_c)^2$ and $N_s(T) = 1-(T/T_c)^4$ are added to mimic the
temperature dependence of $N_s(T)$ for nodal ($d$-wave) and fully
gaped ($s$-wave) superconductivity, respectively. As it has
already been discussed in Ref.~\cite{Chanda}, the former curve
describes the data reasonably well, whereas the fully-gaped
behavior can be ruled out \cite{gap}. Here we added theoretical
results gained from $d$-wave Eliashberg theory calculations for
unitary scattering with $\Gamma^+ = 15\,$meV and Born limit
scattering with $t^+=15\,$meV. Both results agree reasonably well
with experiment over the whole temperature range, thus supporting
nodal superconductivity.

As we measure at a non-zero frequency, the normal electrons always
contribute to $\lambda_{L}$. A comprehensive discussion on this
issue is given in Ref.~\cite{Dordevic}. Obviously, the
normal-electron contribution is minimal at the lowest $T$. Hence,
the lowest-temperature behavior is most relevant for a comparison
between experiment and theory which calculates $\lambda_{L}$ in
the $\omega \rightarrow 0$ limit. To provide this comparison, we
plot the experimental and calculated low-temperature variation of
the normalized penetration depth,
$[\lambda_L(T)-\lambda_L(0)]/\lambda_L(0)$, as a function of the
temperature squared in Figure~\ref{lambda}(b). There is no
question that the results for unitary scattering reproduce best
the data. This result also agrees rather nicely with the findings
reported by Schachinger {\it et al.} \cite{ESchach} for PCCO.

\section{Conclusions}

In our broadband investigation of the optical response of thin PCO
films, we directly observed the formation of the superconducting
condensate at $T < T_{c}$. Eliashberg analysis of the optical
spectra and of the temperature dependence of the penetration depth
proves PCO to be a rather dirty $d$-wave superconductor. In the
superconducting state the scattering off impurities is best
described within the unitary limit. The Eliashberg function
$I^2\chi(\omega)$ develops only marginal changes in passing
through $T_c$. Consequently, the mass enhancement factor $\lambda$
changes very little from 4.16 at $30\,$K and to 4.25 at $4\,$K.
This is a rather big value but one has to keep in mind that
scattering off impurities is pair-breaking in $d$-wave
superconductors and a rather large value of $\lambda$ is required
to ensure a $T_c$ of $27\,$K in our particular case.

\ack

We are very grateful to Dr. Hideki Yamamoto for his work on sample
preparation and for useful discussions.


\begin{thebibliography}{99}
\bibitem{gap} Formally, a very small full gap cannot be excluded,
because we do not measure at temperatures below 2.5 K.
\end{thebibliography}
\end{document}